\documentclass[journal]{IEEEtran}

\usepackage[english]{babel}
\usepackage[utf8]{inputenc}
\usepackage[pdftex]{graphicx}
\usepackage{subfig}
\usepackage{color}
\usepackage{multirow}
\usepackage[normalem]{ulem}
\usepackage{flushend}

\graphicspath{{.},{./figures/}}

\begin{document}

\title{Experimental Interference Robustness Evaluation of IEEE 802.15.4-2015 OQPSK-DSSS and SUN-OFDM Physical Layers}

\author{
    Pere~Tuset-Peir\'o,
    Francisco V\'azquez-Gallego,
    Jonathan Mu\~noz,\\
    Thomas Watteyne,
    Jesus Alonso-Zarate,
    Xavier Vilajosana
    \thanks{Manuscript received Month Date, 2019; revised Month Date, 2019.}
}

\markboth{ArXiv.org Preprint}%
{Tuset \MakeLowercase{\textit{et al.}}: Experimental interference robustness evaluation of IEEE 802.15.4-2015 OQPSK-DSSS and SUN-OFDM physical layers}

\maketitle

\begin{abstract}
In this paper, we experimentally evaluate and compare the robustness against interference of the OQPSK-DSSS (Offset Quadrature Phase Shift Keying - Direct Sequence Spread Spectrum) and the SUN-OFDM (Smart Utility Network - Orthogonal Frequency Division Multiplexing) physical layers, as defined in the IEEE 802.15.4-2015 standard. 
The objective of this study is to provide a comprehensive analysis of the impact different types of interference produce on these modulations, in terms of the resulting PDR (Packet Delivery Ratio) and depending on the length of the packet being transmitted. 
The results show that the SUN-OFDM physical layer provides significant benefits compared to the ubiquitous OQPSK-DSSS in terms of interference robustness, regardless of the interference type and the packet length. 
Overall, this demonstrates the suitability of choosing the SUN-OFDM physical layer when deploying low-power wireless networks in industrial scenarios, specially taking into consideration the possibility of trading-off robustness and spectrum efficiency depending on the application requirements.
\end{abstract}


\IEEEpeerreviewmaketitle

\section{Introduction}
\label{sec:introduction}


The IEEE~802.15.4 standard~\cite{7460875} was first released in May 2003 and defined a physical layer (PHY) and a MAC (Medium Access Control) layer for WPAN (Wireless Personal Area Networks) operating in the sub-GHz (868~MHz in Europe, 915~MHz in America) and the 2.4~GHz worldwide ISM (Industrial, Scientific and Medical) frequency bands~\cite{953229}. 
The PHY layer was built upon the DSSS-OQPSK (Direct Sequence Spread Spectrum - Offset Quadrature Phase Shift Keying) modulation and provided data rates of 20~kbps and 40~kbps in the sub-GHz bands, respectively, and of 250~kbps in the 2.4~GHz band.
At the MAC layer, the standard defined slotted/synchronized and unslotted/unsynchronized operation based on the CSMA/CA (Carrier Sense Multiple Access with Collision Avoidance) channel access mechanism to trade off bandwidth, latency and energy consumption of the devices. 


Thanks to its simplicity and low-cost, over the years the IEEE~802.15.4 standard has become the basis for multiple low-power wireless communications technologies including ZigBee~\cite{zigbee}, ISA100.11a~\cite{ISA100}, WirelessHART~\cite{WirelessHART}, 6TiSCH~\cite{6tisch} and Thread~\cite{Thread}, among others. 
In fact, the adoption of the IEEE~802.15.4 standard by different standardization bodies and technologies has promoted the revision of the standard (i.e.~three times: in 2006, 2011 and 2015) in order to clarify the operation and to add new features to both the PHY and MAC layers. 
For example, the 2015 standard revision adopted the MAC layer proposals defined in the IEEE~802.15.4e-2012~\cite{6185525} amendment.
Among others, this amendment defined the TSCH (Time Slotted Channel Hopping), a channel access mechanism that combines TDMA (Time Division Multiple Access) and FDMA (Frequency Division Multiple Access) to support industrial requirements, including reliable packet delivery (i.e.~99.999\%) in adverse conditions such as multi-path propagation and external interference.


Similarly, the IEEE 802.15.4-2015 standard revision included three new physical layers targeted to SUN (Smart Utility Network) applications~\cite{wisun}, 
SUN-FSK, SUN-OQPSK and SUN-OFDM, as defined in the IEEE~802.15.4g-2012~\cite{6190698} amendment. 
The SUN-FSK and SUN-OQPSK modulations focus on maintaining backwards compatibility with previous standards and commercially available transceivers, whereas the SUN-OFDM is focused on adding robustness and improving spectrum efficiency at the physical layer.
The benefits brought by SUN-OFDM are mainly thanks to the use of parallel transmissions with orthogonal sub-carriers, each transporting a small portion of the information to be transmitted using a narrow-band modulation.
Such approach provides better robustness against multi-path propagation (as sub-channels can be considered flat fading, which are inherently robust to inter-symbol and inter-frame interference) and external interference (as narrow-band interference only affects a portion of the sub-channels and frequency repetition can be used to overcome its effects), as well as improving spectrum efficiency (sub-channels are optimally spaced, allowing to trade-off robustness or data-rate depending on the application requirements).


The consolidation of the IEEE~802.15.4g-2012 amendment into the IEEE~802.15.4-2015 standard, as well as the appearance of the first commercially available radio transceivers supporting it, entails a shift in the design, development and deployment of low-power wireless networks, with special interest in the industrial domain, where the propagation and interference effects have largely slowed or limited the adoption of such technologies.
Taking that into account, in this paper we focus on empirically evaluating the interference robustness of the SUN-OFDM and compare it against OQPSK-DSSS. 
Our goal is to provide evidence of the robustness of the different modulations, including their possible configurations and under different operating conditions, to understand the benefits and limitations when deploying such technologies.
To the best of our knowledge, these results are novel and help researchers and practitioners deploy low-power wireless networks that are more robust and make an efficient use of the spectrum in real-world scenarios.


The remainder of this article is organized as follows.
Section~\ref{sec:overview} provides an overview on the PHYs defined in the IEEE~802.15.4-2015 standard that are evaluated in this article. 
Section~\ref{sec:related} provides a survey of the research related to the SUN-OFDM physical layer for low-power wireless networks.
Section~\ref{sec:methodology_setup} presents the methodology and the setup used to conduct the experiments to evaluate the robustness of the SUN-OFDM and OQPSK-DSSS PHYs. 
Section~\ref{sec:results} presents and summarizes the results obtained from the measurement campaign.
Section~\ref{sec:discussion_and_recommendations} discusses the results obtained and presents a series of recommendations. 
Finally, Section~\ref{sec:conclusions} concludes this article.

\section{IEEE 802.15.4-2015 Overview}
\label{sec:overview}

The 2015 version of the IEEE 802.15.4 standard offers 18~different physical layers, each targeting specific applications and market segments. 
In this section, we present and briefly describe the two physical layers analyzed in this paper, that is, OQPSK-DSSS and SUN-OFDM.

\subsection{IEEE 802.15.4-2015 OQPSK-DSSS}

OQPSK-DSSS was introduced in the original IEEE~802.15.4 standard in 2003. 

OQPSK-DSSS uses 16-ary quasi-orthogonal modulation where each data symbol represents 4~bits of information. 
Therefore, there are 16 possible symbols and each of them is represented by a PRNG (Pseudo-Random Noise Generator) chip sequence that is an almost orthogonal.

When operating in the 780~MHz, 915~MHz, 2380~MHz and 2450~MHz bands, the PRN sequence is 32~chips long for 4~bits of data, hence providing a data rate of 250~kbps. 
In contrast, when operating in the 868~MHz band, the PRN sequence is 16~chips long for 4~bits of data, thus offering a data rate of 100~kbps.

\begin{figure*}[ht!]
    \centering
    \includegraphics[width=0.8\textwidth]{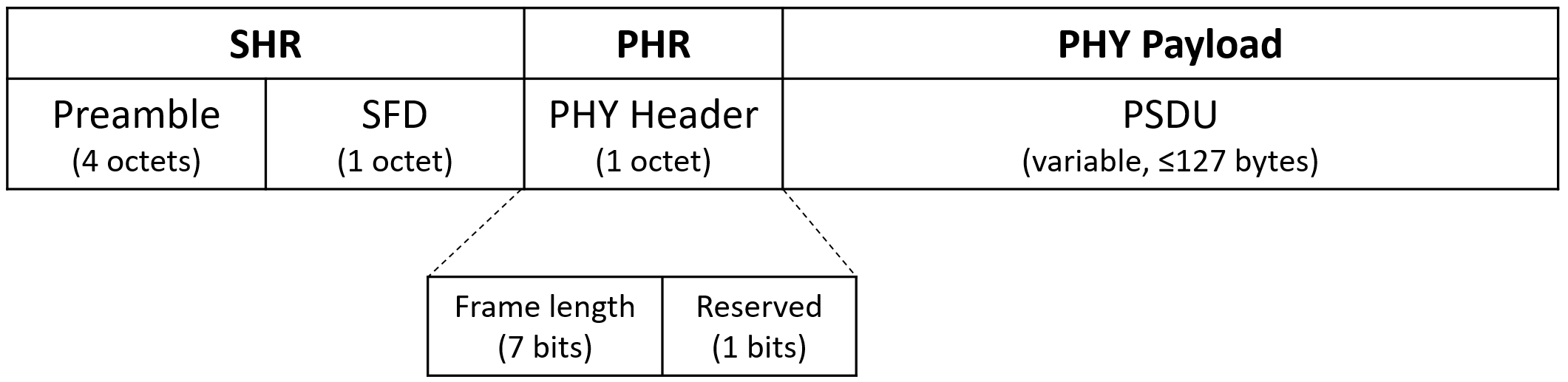}
    \caption{IEEE 802.15.4-2015 OQPSK-DSSS frame format.}
    \label{fig:oqpsk-payload}
\end{figure*}

OQPSK-DSSS frames are composed of a SHR (Synchronization Header), PHR (PHY Header) and PSDU (PHY Service Data Unit), as depicted in Fig.~\ref{fig:oqpsk-payload}. 
The SHR contains a preamble and a SFD (Start of Frame Delimiter); the PHR contains the frame length (7~bits) and a reserved bit.
The maximum length of the PSDU is 127~bytes.

OQPSK-DSSS used in the 2450~MHz band with 250~kbps data rate is the most common PHY configuration deployed when this standard is used.

\subsection{IEEE 802.15.4 SUN-OFDM}

The SUN-OFDM (Orthogonal Frequency Division Multiplexing) physical layer was rolled into the IEEE 802.15.4-2015 version of the standard alongside with SUN-FSK (Frequency Shift Keying) and SUN-OQPSK (Offset Quadrature Phase Shift Keying).
Whereas FSK and OQPSK are two well-known modulation techniques that have been widely used in low-power wireless communications thanks to its simplicity, performance and low-cost, OFDM has been commonly used in other communication domains such as fixed and cellular access (e.g.,~xDSL, WiMAX, LTE), as well as wired and wireless LANs (e.g.,~PLC, Wi-Fi) due to the stringent processing, memory and energy consumption requirements. 
 
The SUN-OFDM PHY can operate in both the sub-GHz bands (400, 700, 800 and 900~MHz) and the 2.4~GHz (2400-2483.5~MHz) with data rates from 50~kbps to 800~kbps depending on the OFDM option and MCS value, as summarized in Table~\ref{tab:sun-ofdm}.

As it can be observed, SUN-OFDM defines 4~options, numbered from 1 to 4, which determine how many sub-carriers are grouped together in order to form an OFDM channel.
Option~1, the largest, groups 104~sub-carriers and occupies a bandwidth of 1094~kHz with 1200~kHz of channel spacing. 
Option~4, the narrowest, groups 14~sub-carriers and occupies a bandwidth of 156~kHz with 200~kHz of channel spacing.
On top of that, the MCS (Modulation and Coding Scheme) parameter, numbered from 0 to 6, determines how each sub-carrier is modulated (available modulation for each carrier are BPSK, QPSK and 16-QAM), whether \textit{frequency repetition} is applied (4x, 2x or no frequency repetition) and the FEC (Forward Error Correction) rate applied to the input data stream for each sub-carrier ($R=1/2$ or $R=3/4$).

Regardless of the OFDM option and MCS value, the sub-carrier spacing and symbol rate is constant: 10416-2/3~Hz and 8-1/3~ksymbols/s.
Each symbol in the OFDM physical layer is 120~$\mu$s long, although the actual symbol transmission time is divided into a BS (Base Symbol) time of 96~$\mu$s and a CP (Cyclic Prefix) time of 24~$\mu$s.
For each symbol, the CP is a replica of the last 1/4 of the BS (24~$\mu$s) and is copied at the front, making the signal periodic.
The long duration of an OFDM symbol and the fact that it is cyclic makes it robust against long spread time environments resulting in multi-path propagation that may cause ISI (Inter-Symbol Interference)~\cite{mccune2013emperor}.

\begin{figure*}[ht!]
    \centering
    \includegraphics[width=0.8\textwidth]{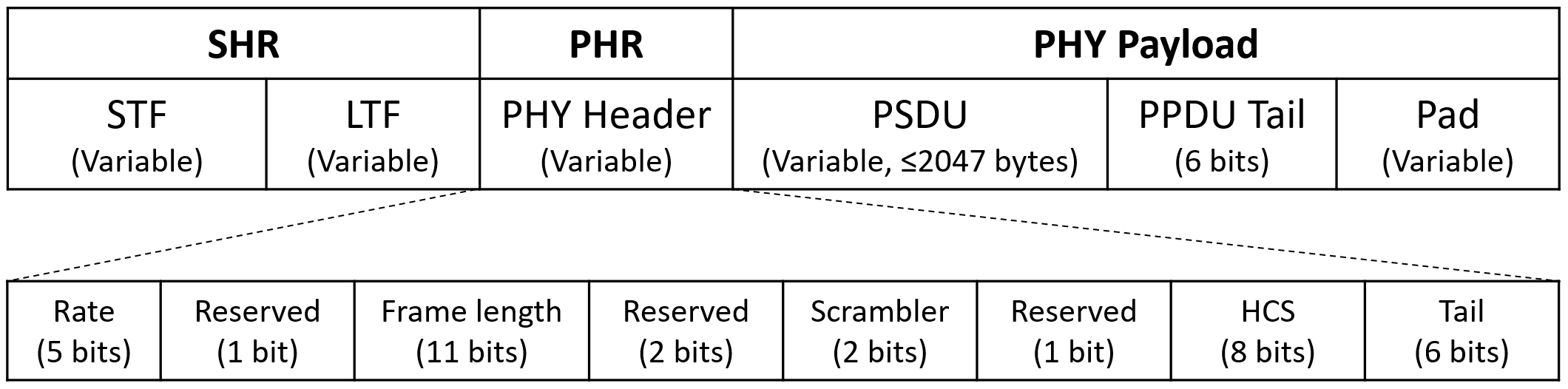}
    \caption{IEEE 802.15.4-2015 SUN-OFDM frame format.}
    \label{fig:ofdm-payload}
\end{figure*}

As per the IEEE 802.15.4-2015 standard, SUN-OFDM frames are composed of a SHR (Synchronization Header), PHR (PHY Header) and PSDU (PHY Service Data Unit), as shown in Fig.~\ref{fig:ofdm-payload}. 
The SHR is composed of a short and long preamble (STF and LTF, resp.), whereas the PHR contains a header with the receiver configuration for the PHY payload (modulation and data rate, frame length, scrambler configuration and a redundancy check).
The OFDM header (SHR and PHR) is transmitted using the lowest supported MCS level for the OFDM option being used. 
That is, for OFDM1 and OFDM2 the header is transmitted using MCS0 (BPSK, R=1/2, 4x repetition);
         for OFDM3 the header is transmitted using MCS1 (BPSK, R=1/2, 2x repetition);
         for OFDM4 the header is transmitted using MCS2 (QPSK, R=1/2, 2x repetition).
Therefore, for any device implementing any of the OFDM options, implementation of the BPSK and QPSK is mandatory and support of 16-QAM is optional.
Finally, the maximum length of the PSDU is 2047~bytes, large enough to transport a complete IPv6 packet without fragmentation.

\begin{table*}[]
    \centering
    \begin{tabular}{|c|c|c|c|c|c|c|c|}
    \hline
    \textbf{Type} & \textbf{Mode} & \textbf{Modulation} & \textbf{Coding rate} & \textbf{Frequency repetition} & \textbf{\begin{tabular}[c]{@{}c@{}}Channel/Nominal\\ bandwidth (kHz)\end{tabular}} & \textbf{\begin{tabular}[c]{@{}c@{}}Total/Data/Pilot\\ tones\end{tabular}} & \textbf{\begin{tabular}[c]{@{}c@{}}Effective datarate\\ (kbps)\end{tabular}} \\ \hline
     & MCS0 & BPSK & 1/2 & 4x &  &  & 100 \\ \cline{2-5} \cline{8-8} 
    \textbf{OFDM1} & MCS1 & BPSK & 1/2 & 2x & 1200 / 1094 & 104/96/8 & 200 \\ \cline{2-5} \cline{8-8} 
     & MCS2 & OQPSK & 1/2 & 2x &  &  & 400 \\ \cline{2-5} \cline{8-8} 
     & MCS3 & OQPSK & 1/2 & 0x &  &  & 800 \\ \hline
     & MCS0 & BPSK & 1/2 & 4x &  &  & 50 \\ \cline{2-5} \cline{8-8} 
     & MCS1 & BPSK & 1/2 & 2x &  &  & 100 \\ \cline{2-5} \cline{8-8} 
    \textbf{OFDM2} & MCS2 & OQPSK & 1/2 & 2x & 800 / 552 & 52/48/4 & 200 \\ \cline{2-5} \cline{8-8} 
     & MCS3 & OQPSK & 1/2 & 0x &  &  & 400 \\ \cline{2-5} \cline{8-8} 
     & MCS4 & OQPSK & 3/4 & 0x &  &  & 600 \\ \cline{2-5} \cline{8-8} 
     & MCS5 & 16-QAM & 1/2 & 0x &  &  & 800 \\ \hline
     & MCS1 & BPSK & 1/2 & 2x &  &  & 50 \\ \cline{2-5} \cline{8-8} 
     & MCS2 & OQPSK & 1/2 & 2x &  &  & 100 \\ \cline{2-5} \cline{8-8} 
    \textbf{OFDM3} & MCS3 & OQPSK & 1/2 & 0x & 400 / 281 & 26/24/2 & 200 \\ \cline{2-5} \cline{8-8} 
     & MCS4 & OQPSK & 3/4 & 0x &  &  & 300 \\ \cline{2-5} \cline{8-8} 
     & MCS5 & 16-QAM & 1/2 & 0x &  &  & 400 \\ \cline{2-5} \cline{8-8} 
     & MCS6 & 16-QAM & 3/4 & 0x &  &  & 600 \\ \hline
     & MCS2 & OQPSK & 1/2 & 2x &  &  & 50 \\ \cline{2-5} \cline{8-8} 
     & MCS3 & OQPSK & 1/2 & 0x &  &  & 100 \\ \cline{2-5} \cline{8-8} 
    \textbf{OFDM4} & MCS4 & OQPSK & 3/4 & 0x & 200 / 156 & 14/12/2 & 150 \\ \cline{2-5} \cline{8-8} 
     & MCS5 & 16-QAM & 1/2 & 0x &  &  & 200 \\ \cline{2-5} \cline{8-8} 
     & MCS6 & 16-QAM & 3/4 & 0x &  &  & 300 \\ \hline
    \end{tabular}%
    \caption{SUN-OFDM parameters.}
    \label{tab:sun-ofdm}
\end{table*}

\section{Related Work}
\label{sec:related}

Despite being introduced in the IEEE~802.15.4g-2012 amendment, the SUN-OFDM PHY has not received a lot of attention from the research community to understand the potential benefits (and pitfalls) of applying this technology in low-power wireless communications for industrial applications.
This is mainly due to the lack of commercially available radio transceivers that support all standard modes.
In this section, we provide a summary of related work that has evaluated the performance of the PHYs proposed in the IEEE~802.15.4g-2012 amendment.
The results are presented in chronological order. 

In~\cite{8539100}, the authors experimentally evaluate various LPWAN (Low-Power Wide-Area Network) technologies for industrial sensing applications, including condition monitoring.
The results, which include LoRa (Long Range) and IEEE~802.15.4g (using SUN-FSK only), conclude that with respect to IEEE~802.15.4g (SUN-FSK), LoRa provides the largest communication range (2x) at the expense of an increased energy consumption (10x).

In~\cite{5936241}, the authors study the coexistence mechanisms of homogeneous and heterogeneous IEEE~802.15.4g systems for SUN, and perform a coexistence analysis to evaluate the performance degradation of a victim system in the presence of an interferer.
They show that with a distance over 30~m the victim and the interferer are able to coexist even without any higher layer mechanism that enforce coexistence.

In~\cite{8376972}, the authors study the coexistence of IEEE~802.11ah and IEEE~802.15.4g networks, which provide communication ranges above 1~km, and are specifically targeted at outdoor IoT (Internet of Things) applications. 
Simulation results show that co-located IEEE~802.11ah networks can severely impact the operation of IEEE~802.15.4g networks due to the higher ED (Energy Detection) threshold and the faster back-off mechanism.
Moreover, the differences in modulation scheme and frame structure limit the capability to implement automatic mechanisms to enforce cooperation. 
Hence, self-coexistence mechanism need to be enforced to mitigate the effects of interference in the network performance. 

More recently, in~\cite{hal-01968648} and~\cite{hal-01718175} the authors evaluate the suitability of the IEEE~802.15.4g-2012 physical layers for environmental and smart building applications respectively, showing that SUN-OFDM is affected by multi-path fading and external interference in a similar way as OQPSK-DSSS. 
Based on these results, in~\cite{hal-01756523}, the authors conclude that the SUN-OFDM physical layer alone does not yield the robustness required for industrial applications and, thus, it recommends to add channel hopping techniques to combat its effects.

\section{Methodology and Setup}
\label{sec:methodology_setup}

In this section, we present the research methodology that we have used to evaluate the interference robustness of the SUN-OFDM modes and compare it with the OQPSK-DSSS mode, both defined in the IEEE~802.15.4-2015 PHYs.

\subsection{Evaluation Methodology}
\label{sec:methodology}

\begin{figure*}[ht!]
    \centering
    \includegraphics[width=0.8\textwidth]{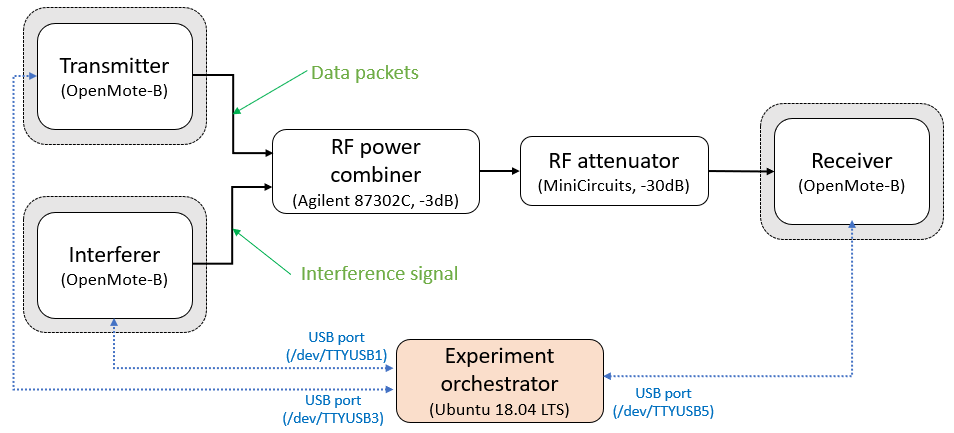}
    \caption{Evaluation setup. Note that the grey box around the transmitter, interferer and receiver devices represents a RF shielding enclosure used to ensure that communication between devices can only occur through the RF coaxial cables.}
    \label{fig:setup}
\end{figure*}

To evaluate the interference robustness of the SUN-OFDM PHY and compare it to the OQPSK-DSSS PHY, we use the setup depicted in Fig.~\ref{fig:setup}.
The setup consists of a transmitter, a receiver and an interferer device connected through an RF power combiner (further described in Section~\ref{sec:setup}). 

For the SUN-OFDM modulation, we select the operating modes that provide a similar data-rate of the OQPSK-DSSS mode, 250~kbps. 
In particular, we select the OFDM1-MCS1, OFDM2-MCS2, OFDM3-MCS3 and OFDM4-MCS5 modes that provide a data-rate of 200~kbps but with different PHY properties (i.e.~channel bandwidth, modulation, coding rate and frequency repetition) as presented in Section~\ref{sec:overview} and summarized in Table~\ref{tab:summary}. 

\begin{table*}[]
    \centering
    \begin{tabular}{|c|c|c|c|c|c|c|c|c|}
    \hline
    \textbf{Name} & \textbf{Mode} & \textbf{Modulation} & \textbf{\begin{tabular}[c]{@{}c@{}}Channel\\ coding\end{tabular}} & \textbf{\begin{tabular}[c]{@{}c@{}}Frequency\\ repetition\end{tabular}} & \textbf{\begin{tabular}[c]{@{}c@{}}Receiver sensitivity\\ (dBm)\end{tabular}} & \textbf{\begin{tabular}[c]{@{}c@{}}Effective data-rate\\ (kbps)\end{tabular}} & \textbf{\begin{tabular}[c]{@{}c@{}}Channel bandwidth\\ (kHz)\end{tabular}} & \textbf{Abbreviation} \\ \hline
    \begin{tabular}[c]{@{}c@{}}OQPSK-\\DSSS\end{tabular} & N/A & OQPSK & N/A & N/A & -103 & 250 & 5000 & OQPSK-DSSS \\ \hline
    \begin{tabular}[c]{@{}c@{}}OFDM\\ Option 1\end{tabular} & MCS1 & BPSK & 1/2 & 2x & -109 & 200 & 1200 & OFDM1-MCS1 \\ \hline
    \begin{tabular}[c]{@{}c@{}}OFDM\\ Option 2\end{tabular} & MCS2 & QPSK & 1/2 & 2x & -108 & 200 & 800 & OFDM2-MCS2 \\ \hline
    \begin{tabular}[c]{@{}c@{}}OFDM\\ Option 3\end{tabular} & MCS3 & QPSK & 1/2 & 0x & -107 & 200 & 400 & OFDM3-MCS3 \\ \hline
    \begin{tabular}[c]{@{}c@{}}OFDM\\ Option 4\end{tabular} & MCS5 & 16-QAM & 1/2 & 0x & -105 & 200 & 200 & OFDM4-MCS5 \\ \hline
    \end{tabular}%
    \caption{IEEE~802.15.4-2015 modulations that have been selected in this study to evaluate their robustness against interference.}
    \label{tab:summary}
\end{table*}

For the measurement campaign, each selected modulation is used as an interferer signal, and tested against all other selected modulations acting as transmitter signals (OFDM1-MCS1, OFDM2-MCS2, OFDM3-MCS3, OFDM4-MCS5, OQPSK-DSSS). 
The procedure is also repeated for two different payload lengths (20~bytes and 120~bytes) in order to understand the effects of packet length on the robustness of each modulation. 
This is relevant as the SUN-OFDM modulation adds symbol redundancy in both time and frequency, whereas OQPSK-DSSS only adds redundancy in terms of the spreading factor gain.
In addition, each base experiment is repeated twice for each transmitter and interferer device (i.e.~exchanging the devices' roles).
We use this approach verify the obtained results and cancel out possible differences in the transmit power between the devices (due to the radio transceiver construction variability and/or the impedance matching of the RF circuit) that may affect the obtained results.

Overall, this gives a total of 10~base experiments, one for each interferer signal and payload length, as presented in Section~\ref{sec:results}.
Each base experiment uses a particular modulation as interferer, and tests the robustness of all the modulations being tested in the measurement campaign, including itself.
For instance, OFDM1-MCS1 is selected as intereferer and tested against OFDM1-MCS1, OFDM2-MCS2, OFDM3-MCS3, OFDM4-MCS5 and OQPSK-DSSS.

To ensure that noise and external interference do not affect the results, we set the transmit power to a value that is well above the sensitivity limit of the radio transceiver for each modulation being tested, and conduct the experiments using RF coaxial cables while each board is enclosed in an individual Faraday cage.
Second, when both transmitter and interferer devices use the same modulation, it is possible that the receiver synchronizes to the header transmitted by the interferer.
To avoid this, the interferer device uses continuous transmit mode and starts transmitting 10~ms before the transmitter device starts sending data packets. 

\subsection{Base experiment}
\label{sec:experiment}

The independent variable of each base experiment is the SINR (Signal to Interference plus Noise Ratio), defined as the ratio between the power of the transmitter signal and the power of the interferer signal.
In contrast, the dependent variable of each experiment is the PDR (Packet Delivery Ratio), defined as the percentage of packets that are successfully received at the receiver under the presence of the interference signal generated by the interferer device. 

For all the experiments, the value of the SINR ranges from -12~dB to +6~dB, meaning that the transmitter power is between 1/16x and 4x the interferer power.
We use such a wide SINR range to ensure that the PDR transitions from 100\% to 0\% regardless of the interferer and transmitter modulation under test. 
In addition, to validate the results obtained of each base experiment, the first test is conducted without interference signal (the interferer device does not transmit during that test).
In that case, the PDR is expected to be 100\%. 
This can be observed with the $No$ mark in the x-axis of the figures presented in Section~\ref{sec:results}.

To determine the PDR for each SINR for a given transmit modulation, the transmitter device sends 1000~frames to the receiver device. 
Frames are transmitted with a 5~ms inter-packet delay.
The payload is composed of the $\{0x00,0x01,...,0x13\}$ sequence for the 20-byte payload and the $\{0x00,0x01,...,0x77\}$ sequence for the 120-byte payload.
Each frames is completed with the FCS field of the PSDU depending on the selected modulation (2~bytes for OQPSK-DSSS, 4~bytes for SUN-OFDM).

During the transmission, the interferer device injects an inteferent signal using the selected modulation for that particular base experiment.
In contrast to data packets, the payload of the interference packets is composed of a 123-byte $\{0x55,...,0x55\}$ sequence.
That payload is transmitted continuously using the continuous transmission feature of the AT86RF215 radio transceiver.
That is, the transceiver operates in packet mode but the packet header (SHR + PHR) is only transmitted once and the payload (PSDU + FCS) is transmitted indefinitely until the experiment is finished. 

There are two conditions that need to be met for a packet to be considered successfully received.
First, the FCS calculated by the receiver radio transceiver based on the received bytes has to match the FCS value attached by the transmitter device at the end of the payload. 
Second, the received bytes have to match the transmitted bytes depending on the packet length for that base experiment.
This ensures that the PDR results are correct even in the unexpected case where the calculated FCS sequence matches the received FCS sequence despite one or multiple bit errors caused by the interferer signal.

\subsection{Setup}
\label{sec:setup}

\begin{figure*}[ht!]
    \centering
    \includegraphics[width=0.8\textwidth]{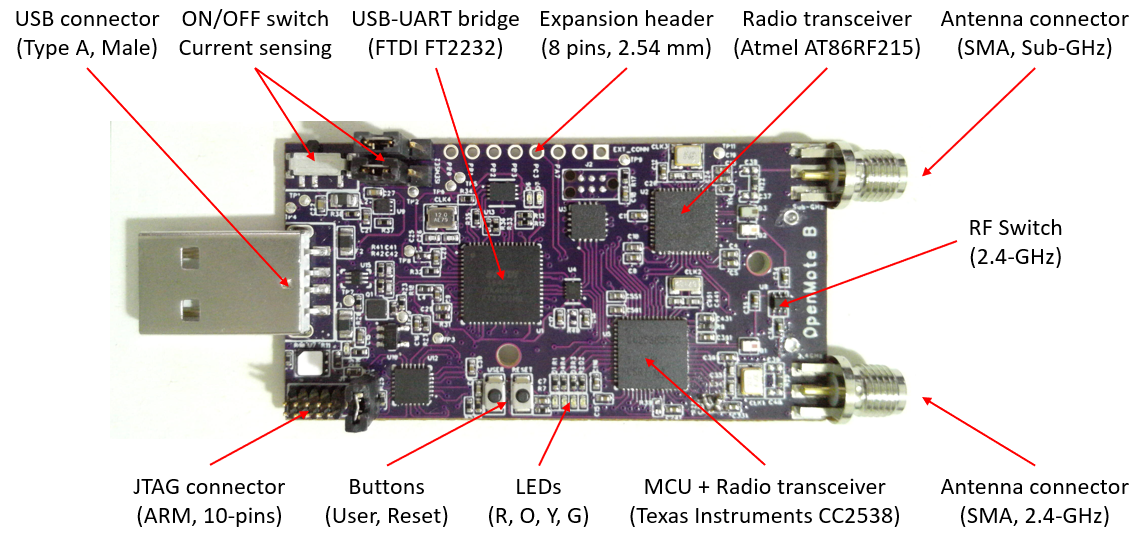}
    \caption{OpenMote-B board. The CC2538 (MCU) and the AT86RF215 (radio transceiver) are interconnected using the SPI bus, whereas the CC2538 (MCU) is connected to the computer using the USB-UART bridge (FTDI FT2232).}
    \label{fig:openmote}
\end{figure*}

For the transmitter, receiver and interferer devices we use the OpenMote-B~\cite{10.1007/978-3-319-25067-0_17, 2893794} shown in Fig.~\ref{fig:openmote}.
The OpenMote-B is equipped with a Texas Instruments CC2538 SoC (System on Chip) and a Microchip AT86RF215 radio transceiver. 
The CC2538~\cite{ti:cc2538} includes an ARM Cortex-M3 micro-controller (32~MHz, 32~kB RAM, 512~kB flash) and a radio transceiver compatible with the IEEE~802.15.4-2006 standard. 
The AT86RF215~\cite{atmel:at86rf215} is a dual-band (sub-GHz and 2.4~GHz) radio transceiver compatible with the IEEE~802.15.4-2015 standard, which supports all Multi-Rate PHY options defined in the IEEE~802.15.4g-2012 amendment for Smart Utility Networks (SUN).

As depicted in Fig.~\ref{fig:setup}, the boards are connected using an Agilent 87302C RF power combiner that operates in the 0.5-26.5~GHz band and provides an input attenuation of 3~dB for the input channels towards the output channel, and a return loss of 24~dB. 
The output of the RF power combiner is further attenuated using a Minicircuits 30~dB RF attenuator to ensure that the received signal at the receiver device is within the receive power limits of the AT86RF215 radio transceiver (-5~dBm according to the data-sheet).

To coordinate the actions between the different devices, a computer running Ubuntu 18.04 LTS is used.
The computer is connected to the OpenMote-B boards over USB, and executes a Python script that orchestrates an experiment.
For each board, the script uploads the radio configuration to be used (i.e.~modulation type, transmit power and packet length, among others) and starts transmission/reception. 
The OpenMote-B boards execute a custom firmware that receives commands over UART (Universal Asynchronous Receiver Transmitter), execute the action commanded (i.e.~transmit a packet, start receiving, etc.) and return the result (i.e.~packet successfully received or not).

\subsection{Radio Calibration}
\label{sec:calibration}

The PDR is evaluated with respect to SINR between the transmit power of the data and the interferer signals.
According to the AT86RF215 radio transceiver datasheet (specifically Fig.~11-2), there is a 6~dB offset between the value of the transmit power configuration register ({\tt RFn\_PAX.TXPWR}) and the actual transmit power of the radio transceiver depending on the selected modulation (OQPSK-DSSS and SUN-OFDM). 

Such difference in transmit power between different modulations can have an impact on the obtained results. 
For instance, an uncontrolled higher transmit power of the data signal results in a better PDR result for that particular modulation.
Similarly, an uncontrolled lower transmit power of the interferer signal also results in a better PDR for that particular modulation.
To validate the offset and obtain the configuration register values that ensure the same transmit power for both modulations, we calibrate using a Rigol DSA-815 spectrum analyzer as the Test Equipment (TE), and an OpenMote-B as the Device Under Test (DUT).

The DUT is connected to TE using a RF cable and the radio calibration procedure is a performed as follows.
For each modulation, the DUT is configured to transmit with a known power configuration register and the TE is configured to measure the power in the band taking into consideration the parameters of the modulation under test (i.e.~center frequency, occupied bandwidth and span). 
The procedure is repeated for all the values of the power configuration register (from 0 to 30 with a 3 unit step) to obtain the relation between those values and the actual transmit powers that are generated by the radio transceiver, as shown in Table~\ref{tab:calibration}.

\begin{table*}[]
    \centering
    \begin{tabular}{|c|c|c|c|c|c|}
    \hline
     & \multicolumn{5}{c|}{\textbf{Transmit power (dBm)}} \\ \hline
    \multicolumn{1}{|c|}{\textbf{RFn\_PAX.TXPWR}} & \textbf{OQPSK-DSSS} & \textbf{OFDM1-MCS1} & \textbf{OFDM2-MCS2} & \textbf{OFDM3-MCS3} & \textbf{OFDM4-MCS54} \\ \hline
    \multicolumn{1}{|c|}{\textbf{30}} & 15 & 9 & 9 & 9 & 9 \\ \hline
    \multicolumn{1}{|c|}{\textbf{27}} & 14 & 8 & 8 & 8 & 8 \\ \hline
    \multicolumn{1}{|c|}{\textbf{24}} & 12 & 5 & 5 & 5 & 5 \\ \hline
    \multicolumn{1}{|c|}{\textbf{21}} & 9 & 2 & 2 & 2 & 2 \\ \hline
    \multicolumn{1}{|c|}{\textbf{18}} & 6 & -1 & -1 & -1 & -1 \\ \hline
    \multicolumn{1}{|c|}{\textbf{15}} & 3 & -3 & -3 & -3 & -3 \\ \hline
    \multicolumn{1}{|c|}{\textbf{12}} & 0 & -7 & -7 & -7 & -7 \\ \hline
    \multicolumn{1}{|c|}{\textbf{9}} & -3 & -11 & -11 & -11 & -11 \\ \hline
    \multicolumn{1}{|c|}{\textbf{6}} & -6 & -14 & -14 & -14 & -14 \\ \hline
    \multicolumn{1}{|c|}{\textbf{3}} & -9 & -16 & -16 & -16 & -16 \\ \hline
    \multicolumn{1}{|c|}{\textbf{0}} & -12 & -18 & -18 & -18 & -18 \\ \hline
    \end{tabular}
    \caption{
        Transmit power values for the OQPSK-DSSS and the SUN-OFDM modulations.
        Notice that the results have been rounded to the closest dB given the resolution of the DUT and the uncertainty of the TE.
    }
    \label{tab:calibration}
\end{table*}

Overall, results are in accordance with the data provided in the datasheet and allow to empirically determine the transmit power configuration register values for the OQPSK-DSSS and the SUN-OFDM modulations.
In particular, we have decided to use a transmit power between -9~dBm and +9~dBm, as these are the values that are common to both modulations.
We then define the default transmit power of +3~dBm, and let the interferer power range from -9~dBm to +9~dBm, corresponding to the SINR between -12~dB and +6~dB described earlier.

\section{Results}
\label{sec:results}

\begin{figure*}[ht!]
    \centering
    \subfloat[PSDU=$20$~bytes\label{}]{{\includegraphics[width=0.45\textwidth]{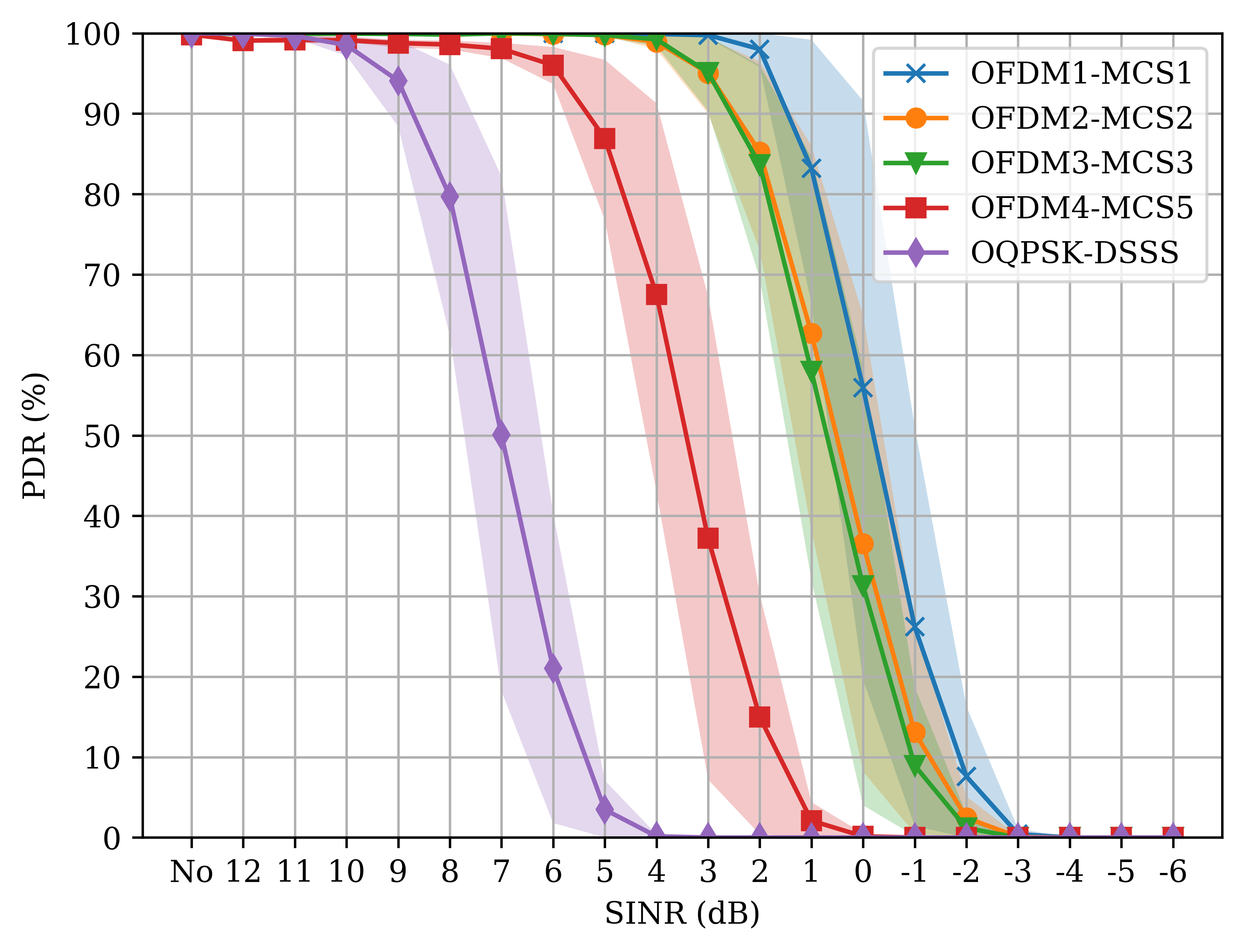}}}
    \qquad
    \subfloat[PSDU=$120$~bytes\label{}]{{\includegraphics[width=0.45\textwidth]{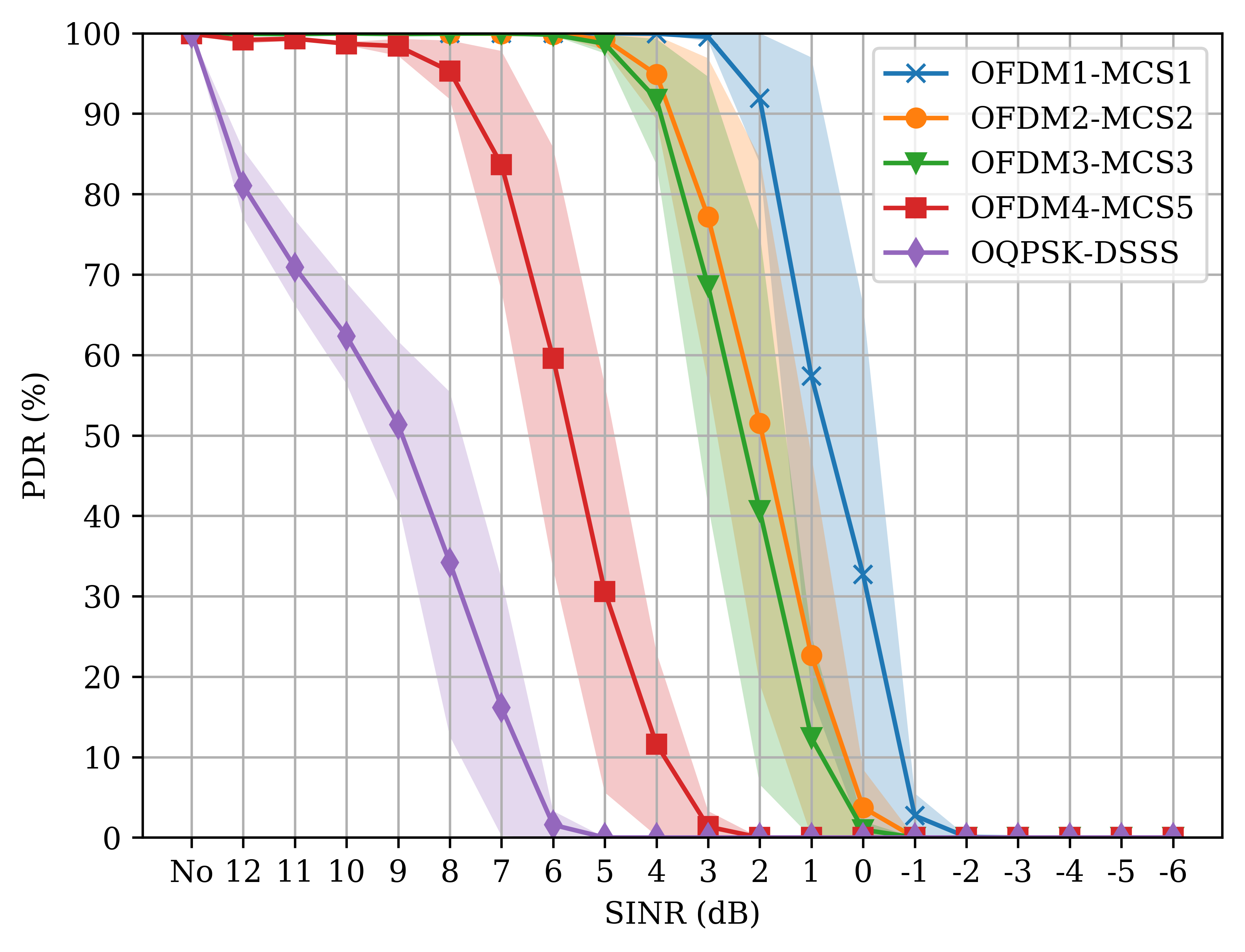}}}
    \caption{OFDM1-MCS1 interference results.}
    \label{fig:ofdm1-mcs1}
\end{figure*}

\begin{figure*}[ht!]
    \centering
    \subfloat[PSDU=$20$~bytes\label{}]{{\includegraphics[width=0.45\textwidth]{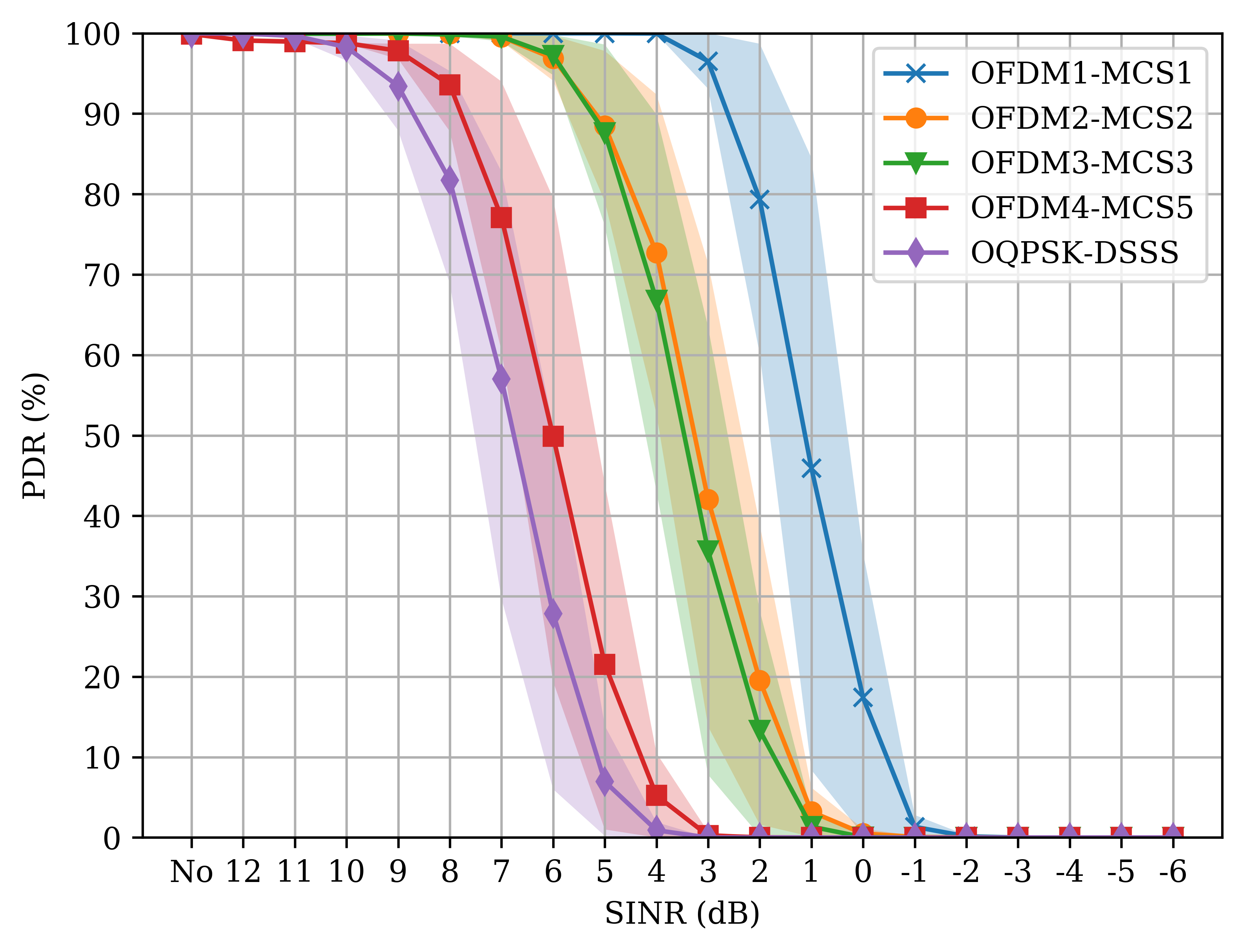}}}
    \qquad
    \subfloat[PSDU=$120$~bytes\label{}]{{\includegraphics[width=0.45\textwidth]{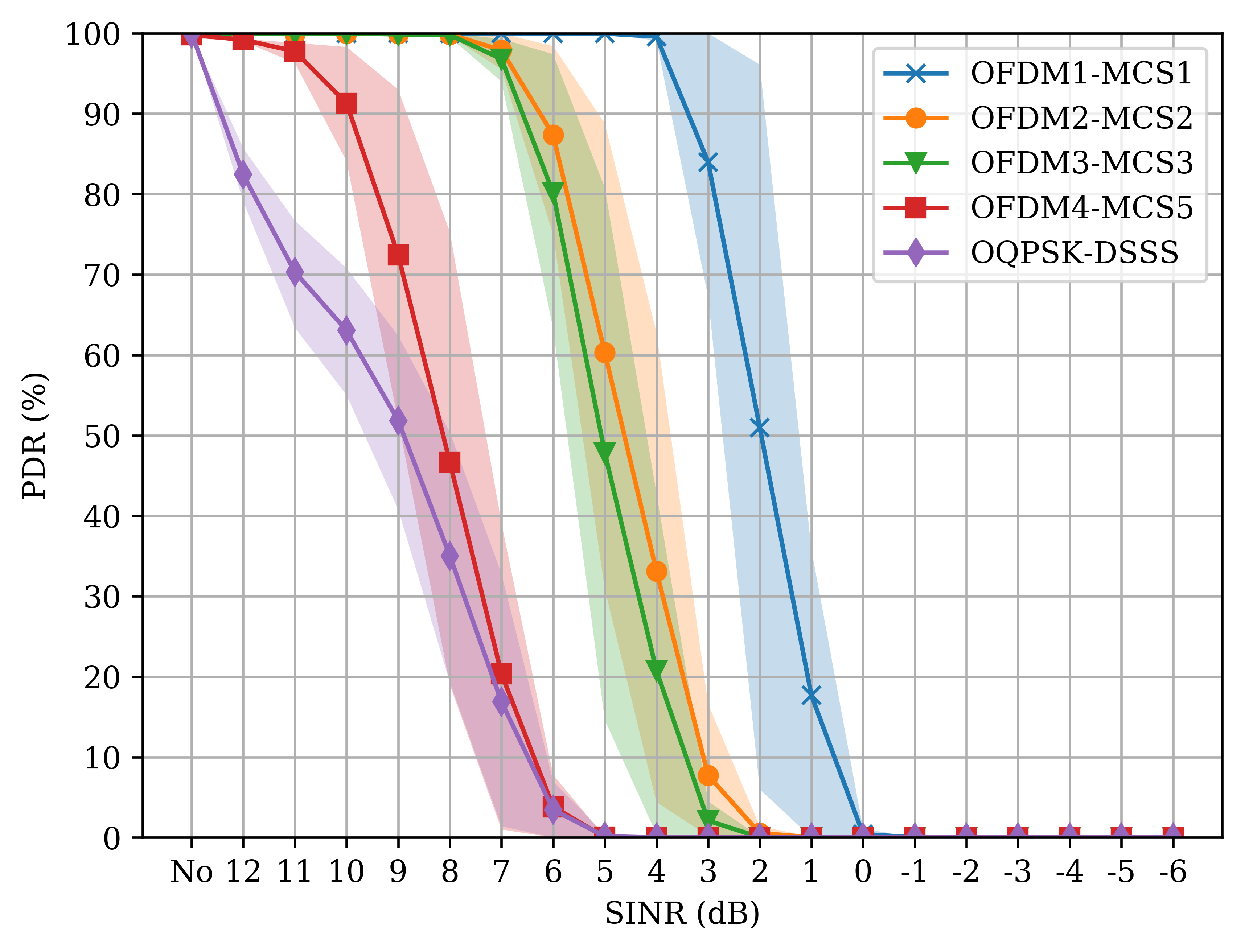}}}
    \caption{OFDM2-MCS2 interference results.}
    \label{fig:ofdm2-mcs2}
\end{figure*}

\begin{figure*}[ht!]
    \centering
    \subfloat[PSDU=$20$~bytes\label{}]{{\includegraphics[width=0.45\textwidth]{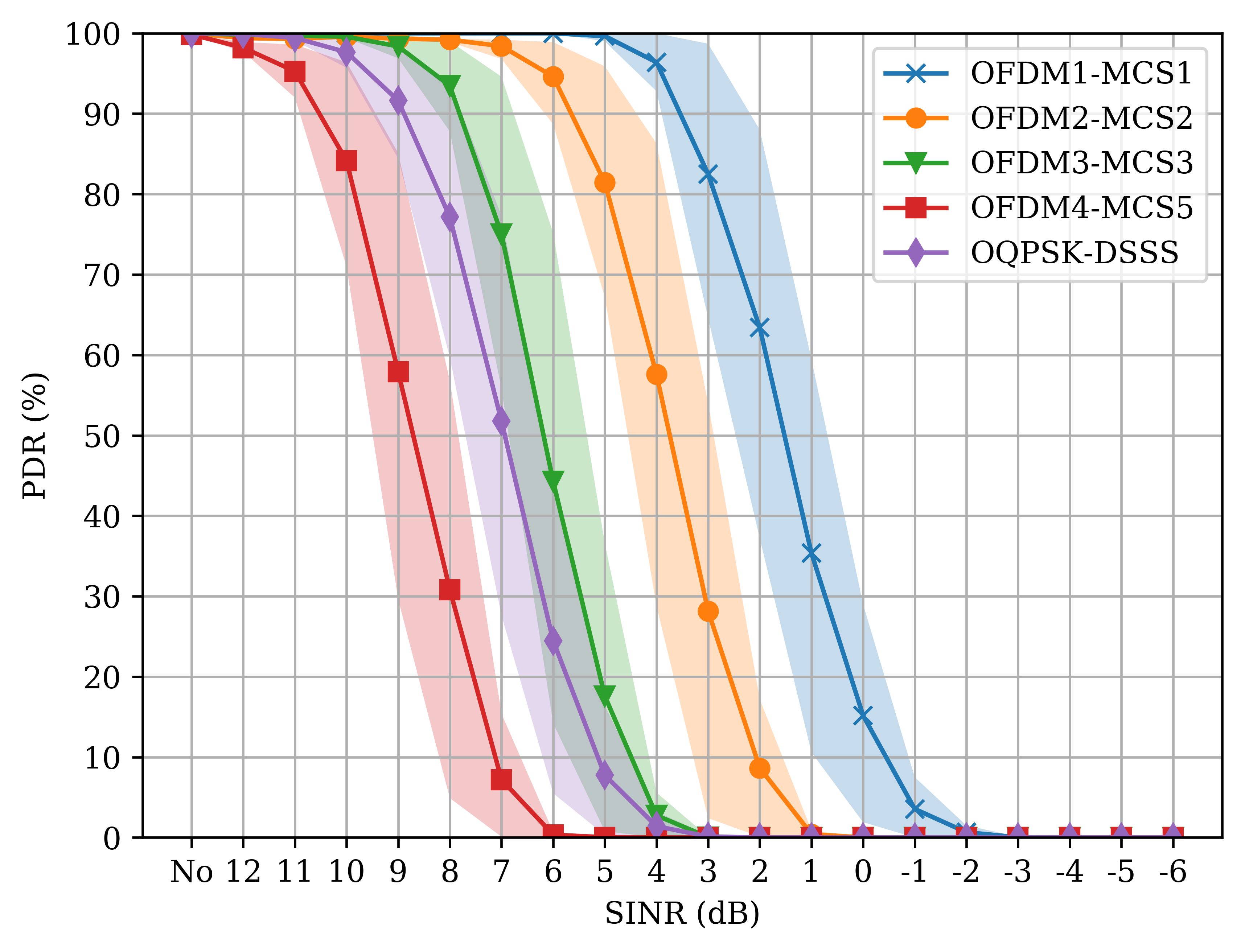}}}%
    \qquad
    \subfloat[PSDU=$120$~bytes\label{}]{{\includegraphics[width=0.45\textwidth]{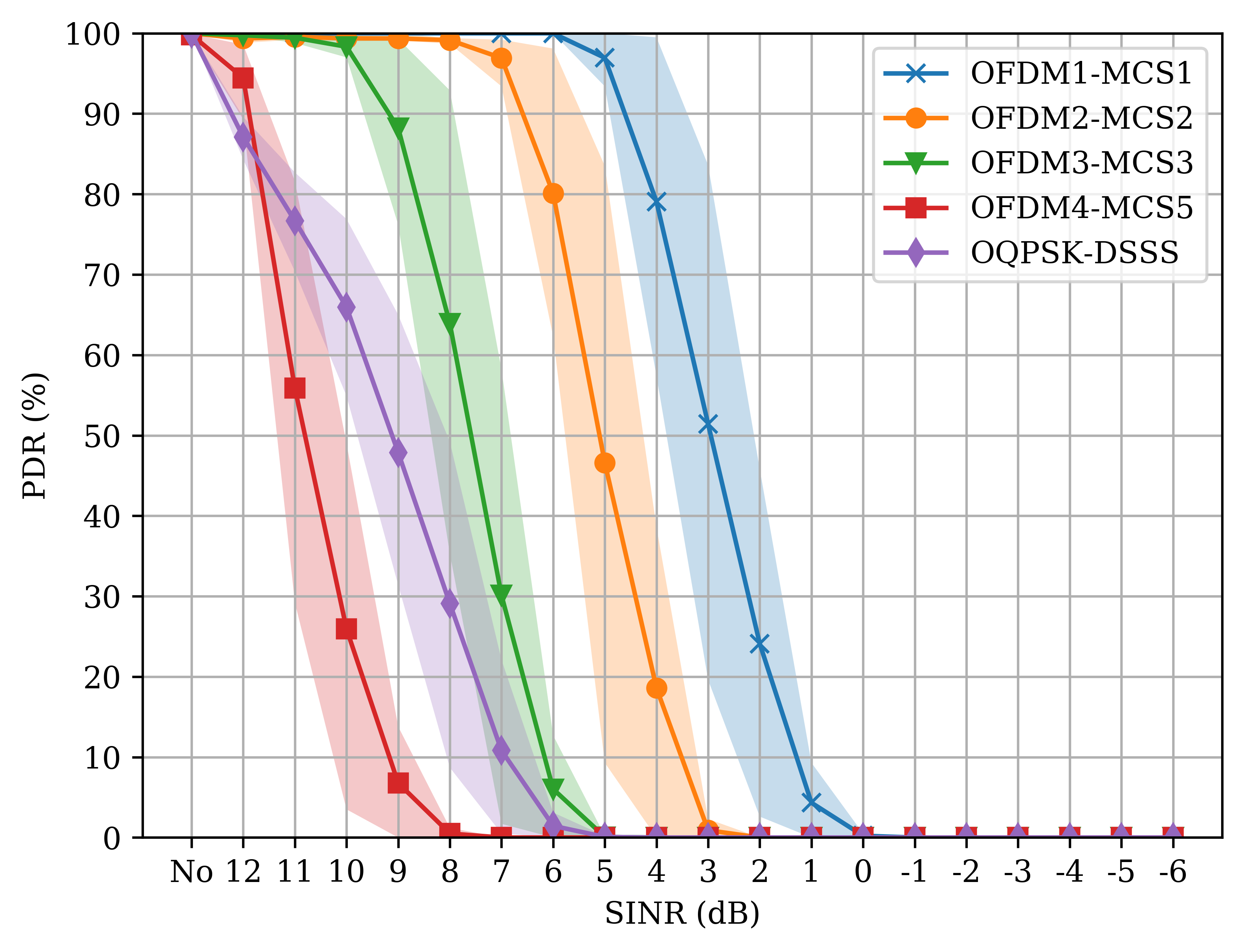}}}%
    \caption{OFDM3-MCS3 interference results.}
    \label{fig:ofdm3-mcs3}
\end{figure*}

\begin{figure*}[ht!]
    \centering
    \subfloat[PSDU=$20$~bytes\label{}]{{\includegraphics[width=0.45\textwidth]{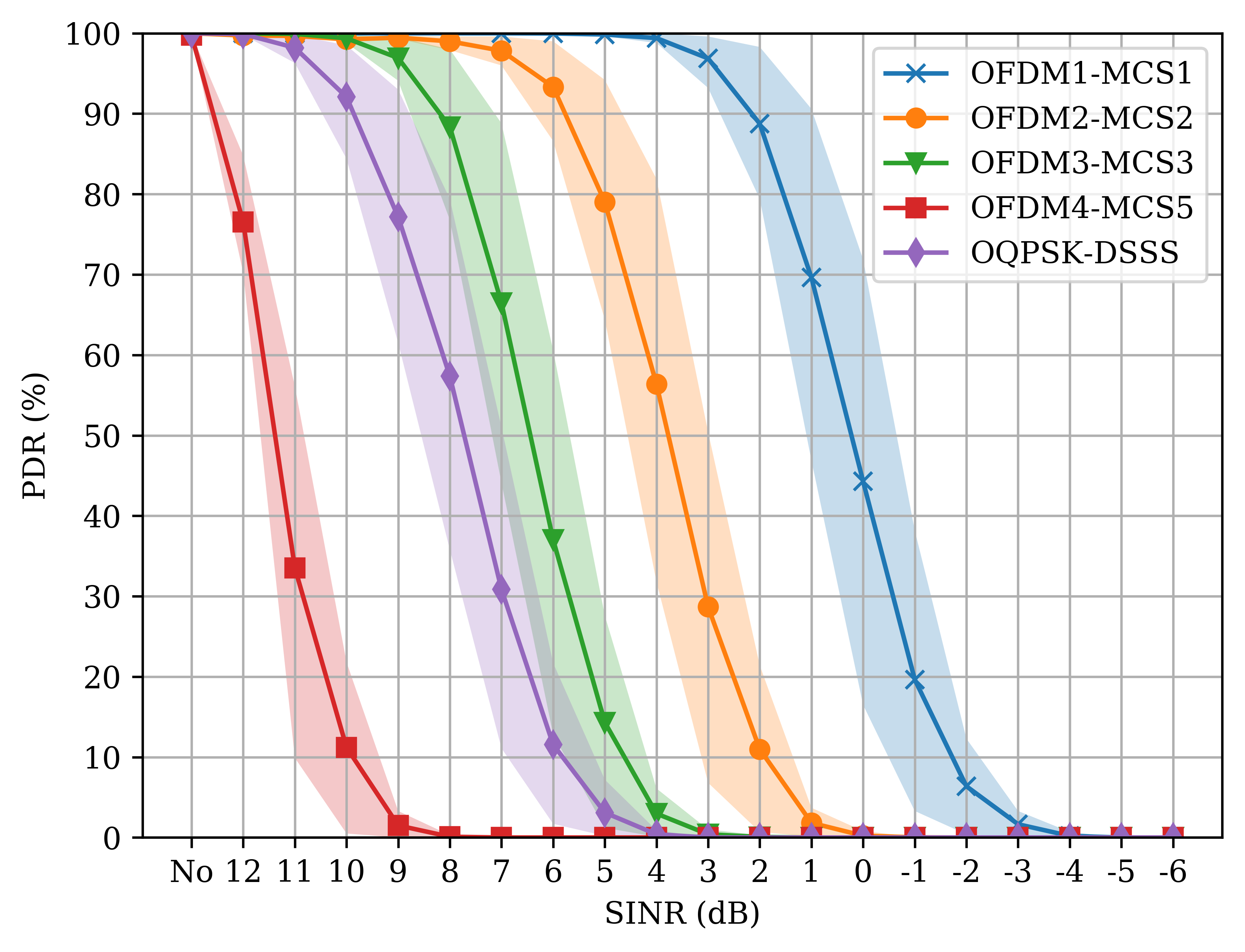}}}%
    \qquad
    \subfloat[PSDU=$120$~bytes\label{}]{{\includegraphics[width=0.45\textwidth]{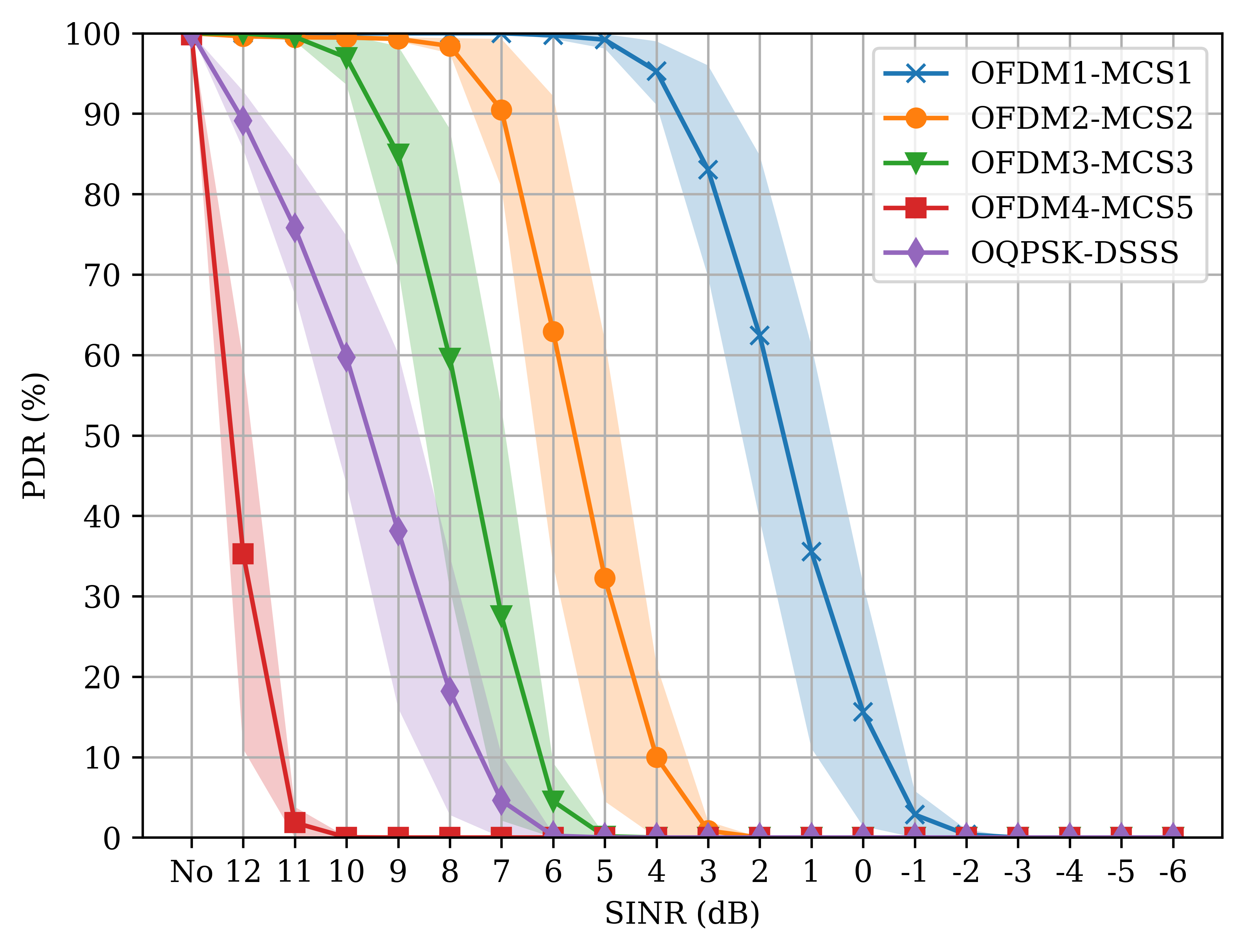}}}%
    \caption{OFDM4-MCS5 interference results.}%
    \label{fig:ofdm4-mcs5}
\end{figure*}

\begin{figure*}[ht!]
    \centering
    \subfloat[PSDU=$20$~bytes\label{}]{{\includegraphics[width=0.45\textwidth]{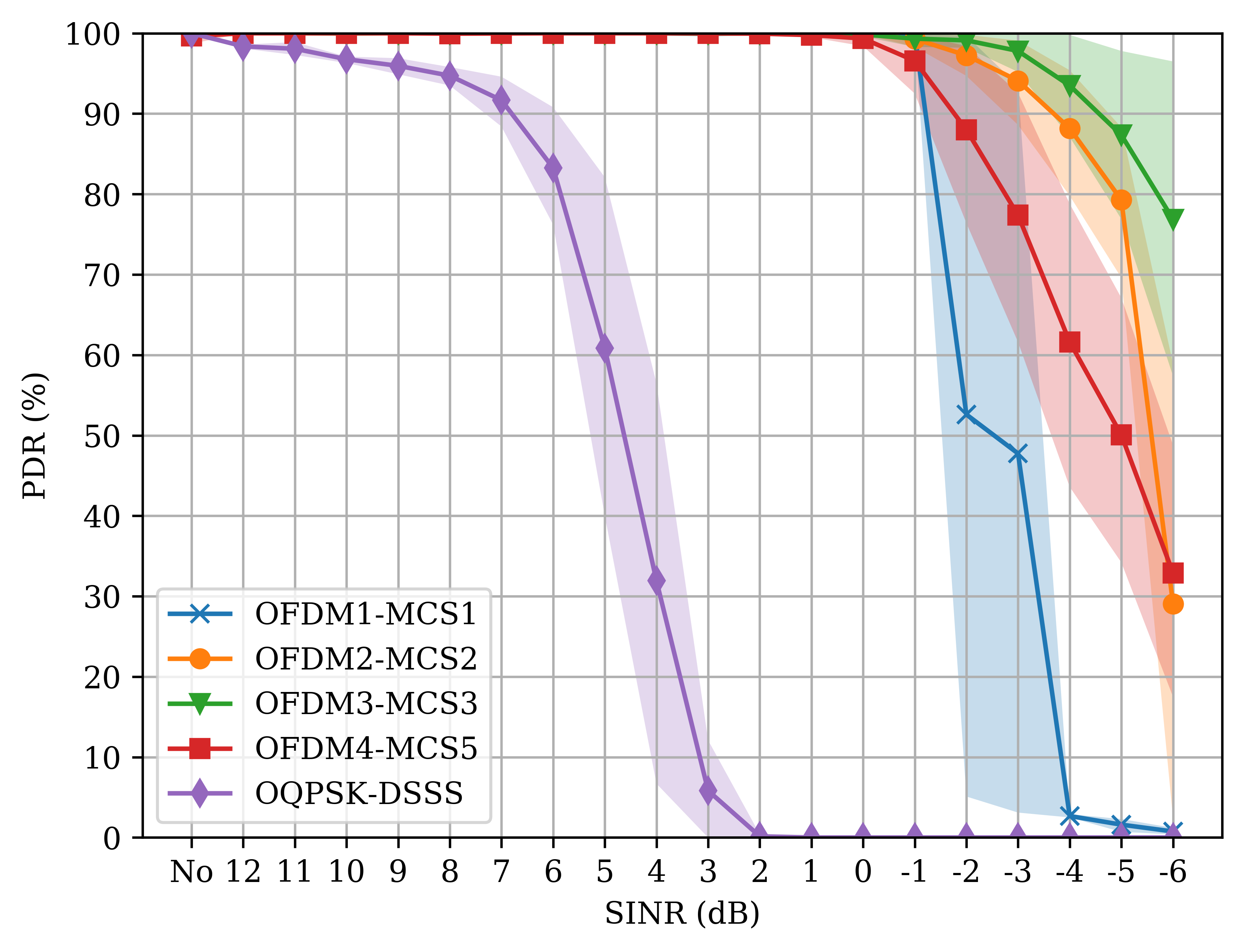}}}
    \qquad
    \subfloat[PSDU=$120$~bytes\label{}]{{\includegraphics[width=0.45\textwidth]{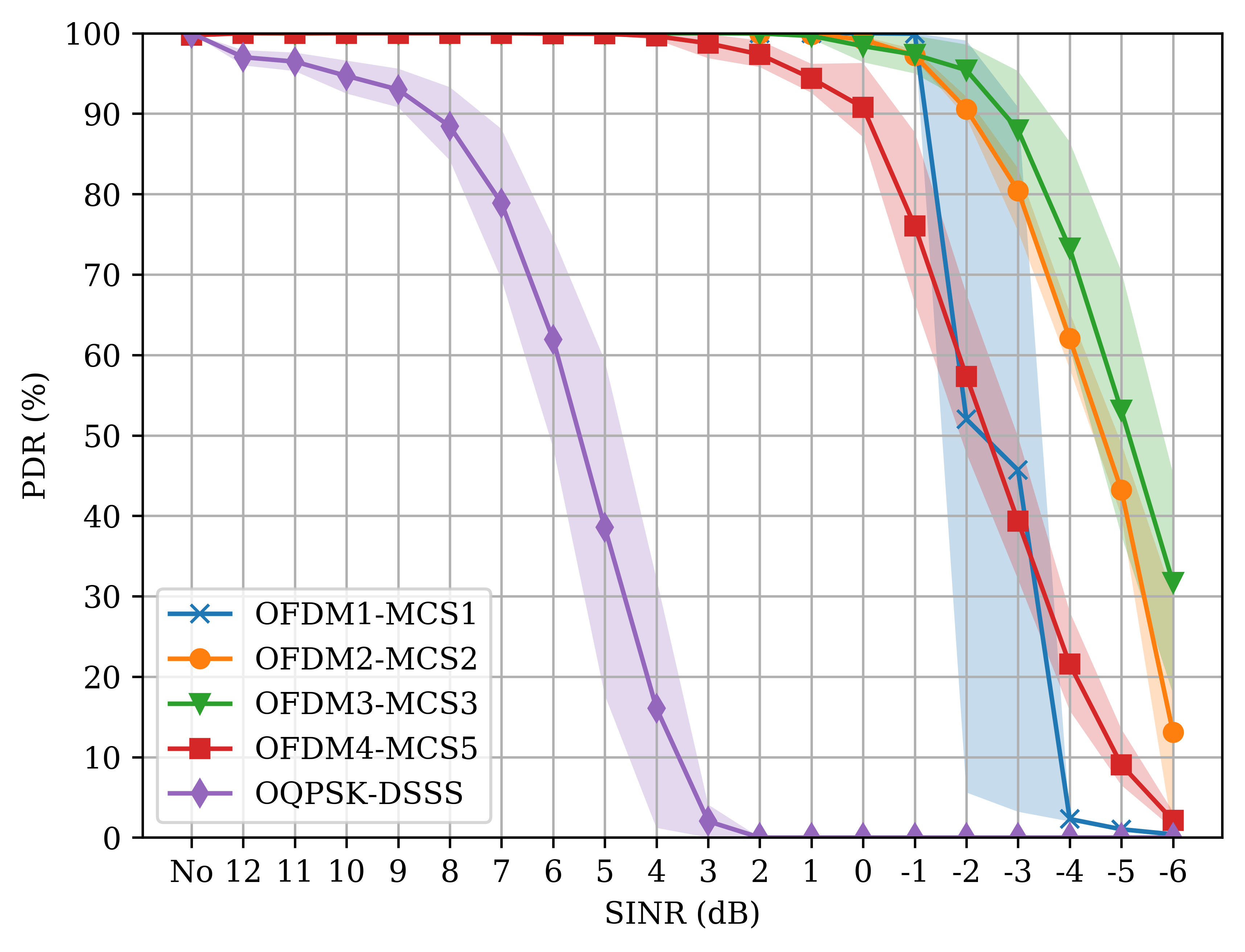}}}
    \caption{OQPSK-DSSS interference results.}
    \label{fig:oqpsk-dsss}
\end{figure*}

Results obtained using the research methodology and the experiment setup described in the previous section are presented for each interferer modulation (OFDM1-MCS1, OFDM2-MCS2, OFDM3-MCS3, OFDM4-MCS5 and OQPSK-DSSS) and packet length (20~bytes and 120~bytes) in Fig.~\ref{fig:ofdm1-mcs1}, \ref{fig:ofdm2-mcs2}, \ref{fig:ofdm3-mcs3}, \ref{fig:ofdm4-mcs5} and~\ref{fig:oqpsk-dsss}, respectively. 
Table~\ref{tab:results} summarizes the required SINR (dB) to ensure PDR$>80$\% for each modulation and packet length.

\begin{table*}[ht!]
    \centering
    \begin{tabular}{|c|c|c|c|c|c|c|c|c|c|c|}
    \hline
     & \multicolumn{10}{c|}{\textbf{Interference type and packet length (bytes)}} \\ \cline{2-11} 
     & \multicolumn{2}{c|}{\textbf{OFDM1-MCS1}} & \multicolumn{2}{c|}{\textbf{OFDM2-MCS2}} & \multicolumn{2}{c|}{\textbf{OFDM3-MCS3}} & \multicolumn{2}{c|}{\textbf{OFDM4-MCS5}} & \multicolumn{2}{c|}{\textbf{OQPSK-DSSS}} \\ \cline{2-11} 
     & \textbf{20} & \textbf{120} & \textbf{20} & \textbf{120} & \textbf{20} & \textbf{120} & \textbf{20} & \textbf{120} & \textbf{20} & \textbf{120} \\ \hline
    \textbf{OQDM1-MCS1} & 1 & 2 & 2 & 3 & 3 & 4 & 2 & 3 & -1 & -1 \\ \hline
    \textbf{OQDM1-MCS2} & 2 & 4 & 5 & 6 & 5 & 6 & 5 & 7 & -5 & -3 \\ \hline
    \textbf{OQDM3-MCS3} & 2 & 4 & 5 & 6 & 8 & 9 & 8 & 9 & -5 & -3 \\ \hline
    \textbf{OQDM4-MCS5} & 5 & 7 & 8 & 10 & 10 & 12 & \textgreater{}12 & \textgreater{}12 & -2 & 0 \\ \hline
    \textbf{OQPSK-DSSS} & 8 & 12 & 8 & 12 & 8 & 12 & 9 & 12 & 6 & 8 \\ \hline
    \end{tabular}%
    \caption{Results summary. Each value in the table is the required SINR (dB) to ensure a PDR$>80$\% for each modulation and packet length.}
    \label{tab:results}
\end{table*}

For example, if we consider an interfering signal modulated using OFDM1-MCS1, we can observe that, for a 20~B PSDU, the OFDM1-MCS1, OFDM2-MCS2 and OFDM-MCS3 physical layers perform similarly, requiring SINR$>$1~dB for PDR$>$80\%.
OFDM4-MCS5 requires SINR$>$5~dB to provide a PDR$>$80\%.
In contrast, for a 120~B PSDU,
    OFDM1-MCS1 requires a SINR$>$2~dB,
    OFDM2-MCS2 and OFDM3-MCS3 require SINR$>$4~dB, and
    OFDM4-MCS5 requires a SINR$>$7~dB.
Finally, for a 20~B and 120~B PSDU, OQPSK-DSSS requires a SINR$>$8~dB and a SINR$>$12~dB for PDR$>$80\%, resp.

These results show that OQPSK-DSSS requires 6~dB higher SINR to achieve PDR$>$80\% when compared to OFDM1-MCS1 under the same interference. 
Considering that OQPSK-DSSS uses a 5~MHz channel bandwidth while OFDM1-MCS1 uses 1.2~MHz channel bandwidth, we can derive that OFDM provides increased spectral efficiency and robustness. 
Similar results can be derived for other combinations, as presented in Table~\ref{tab:results}. 

Interestingly, considering an interfering signal modulated using OQPSK-DSSS, if we consider a system using OFDM1-MCS1 and we compare it to the same system using OQPSK-DSSS, we can observe that OFDM1-MCS1 can still operate with PDR$>$80\% even if the interference level is 1~dB above the signal level. 
This is particularly interesting because OQPSK-DSSS is widely used. 

\section{Discussion and Recommendations}
\label{sec:discussion_and_recommendations}

This section discusses the results outlined in the previous section and presents recommendations to researchers and practitioners, serving as a tool to understand the behavior of each particular modulation under interference conditions, and motivating the use of the SUN-OFDM for deploying low-power wireless systems in industrial scenarios. 

\subsection{Discussion}
\label{sec:discussion}

Following the results presented in Table~\ref{tab:results}. 

First, considering OFDM1-MCS1, we observe that a reduction in the bandwidth of the interference signal (i.e.~going from OFDM2-MCS2 to OFDM4-MCS5 interference) requires a 2~dB SINR increase to provide a PDR$>$80\%. 
This can be attributed to the fact that the reduction in the bandwidth causes an increase in the PSD (Power Spectral Density) of the interference (i.e.~a 3~dB increase each time the bandwidth is reduced by half) meaning that the probability of a symbol being corrupted increases.

Second, the SINR required for OFDM2-MCS2 and OFDM3-MCS3 is equivalent to OFDM1-MCS1 and OFDM2-MCS2 regardless of packet size (2-4~dB and 5-6~dB, resp.). 
However, the effect of OFDM3-MCS3 interference causes a 3~dB SINR increase to the OFDM3-MCS3 signal (i.e.~7-8~dB) to maintain PDR$>$80\%, whereas the OFDM2-MCS2 signal remains unaffected (5-6~dB).
Such results can be attributed to the fact that, despite both OFDM2-MCS2 and OFDM3-MCS3 using the same modulation (OQPSK with coding rate $R=1/2$), the former takes advantage of the 2x frequency repetition providing an additional protection of 3~dB. 
The result is independent of packet length, which confirms the importance of using such protection mechanisms at the physical layer. 

Third, the SINR required for OFDM4-MCS5 to maintain a PDR$>$80\% is higher compared to the other SUN-OFDM modulations and increases as the bandwidth of the interferer is reduced (going from OFDM1-MCS1 to OFDM4-MCS5 interference). 
This is an expected result, as the PSD of the interference increases by 3~dB every time the bandwidth is reduced by two, and taking into account that OFDM4-MCS5 uses an aggressive modulation (16-QAM) in combination with a coding rate $R=1/2$, but does not employ time or frequency symbol repetition, leading to a higher BER (Bit Error Rate) in the presence of interference.

Fourth, the SINR required for the OQPSK-DSSS modulation remains constant (8-12~dB depending on packet length) regardless of the type of the SUN-OFDM interference and, hence, the occupied bandwidth. 
In contrast, OQPSK-DSSS modulation requires a smaller SINR (6-8~dB depending on packet length) when the interference is OQPSK-DSSS.
This can be attributed to a combination of facts including that SUN-OFDM interference is composed of multiple pilots spread over the channel and has a higher PSD, as well as the fact that OQPSK-DSSS  does not provide any additional mechanism to enhance the robustness of the transmission against interference except for the processing gain of DSSS ($10 \cdot log10(2M/250k) = 9$~dB). 

Fifth, SUN-OFDM modulations are robust against OQPSK-DSSS interference, requiring negative SINR values (SINR$<0$~dB) to maintain a PDR$>$80\%. 
This is an expected result that can be attributed to the fact that SUN-OFDM modulation is a combination of narrowband modulation, whereas the OQPSK-DSSS is inherently wideband, i.e.~it has a very low PSD.
In fact, the results show that OFDM2-MCS2 and OFDM3-MCS3 require the same SINR, which is lower than OFDM1-MCS1. 
This can be explained given that OFDM2-MCS2 employs 2x frequency repetition whereas OFDM3-MCS3 occupies half the bandwidth (400~kHz vs. 800~kHz, resulting in a 3~dB PSD increase). 

As a numeric example of the above, for OFDM2-MCS2 the PSD for a transmit power of 0~dBm and a bandwidth of 800~kHz is -29.03~dBm/Hz.
In contrast, for OFDM3-MCS3, the PSD for a transmit power of 0~dBm and a bandwidth of 400~kHz is -26.02~dBm/Hz. Finally, OQPSK-DSSS interference with a transmit power of 0~dBm and a bandwidth of 2~MHz has a PSD of -33.01~dBm/Hz. 
Notice that, for an SINR=0~dB, OFDM2-MCS2 has SINR=4~dB ($-29.03 - -33.01 = 3.98$~dB) whereas OFDM3-MCS3 has SINR=7~dB ($-26.02 - 33.01 = 6.99$~dB). 
In conclusion, when dealing with wideband interference (here, OQPSK-DSSS) using 2x repetition provides a theoretical 3~dB advantage, but that advantage is overcome by the bandwidth reduction that provides a higher PSD. 


Another interesting result comparing the performance of the SUN-OFDM and OQPSK-DSSS physical layers against interference is that the degradation in terms of PDR caused by the increment of the packet length (going from 20~bytes to 120~bytes) is not equal for both modulation types.
For SUN-OFDM, the SINR increase to maintain a PDR$>$80\% is bounded to 2~dB, whereas for OQPSK-DSSS the SINR increase is larger than 3~dB. 
This is thanks to time and frequency diversity, which allows to recover random errors caused by multi-path propagation and external interference.


The SINR spread between operating in ideal conditions (PDR$>$80\%) and degraded conditions (PDR$<$20\%) for all SUN-OFDM modulations requires SINR=3~dB regardless of the packet length. 
In contrast, the SINR spread between operating in ideal conditions and degraded conditions for OQPSK-DSSS depends on packet size and interference type. 
When dealing with SUN-OFDM interference, transitions requires SINR=2~dB for 20-byte packets, and SINR=5~dB for 120-byte packets. 
In contrast, when dealing with OQPSK-DSSS interference, the transition requires SINR=2~dB regardless of packet length.


Comparing OFDM1-MCS1 to OQPSK-DSSS, we observe that for a PDR$>$80\%, it provides an advantage between 5-10~dB, depending on interference type and packet length.
In particular, it provides an advantage of 5~dB for OFDM3-MCS3 interference with packet length of 20~bytes, and an advantage of 10~dB for OFDM1-MCS1 interference with packet length of 120~bytes. 
This is in contrast with the fact that OFDM1-MCS1 uses a channel separation of 1.2~MHz, whereas OQPSK-DSSS uses a channel separation of 5~MHz. 
That is, the spectral efficiency of OFDM1-MCS1 is 0.166~bits/Hz (or 200~kbps/1200~kHz) whereas the spectral efficiency of OQPSK-DSSS is 0.05~bits/Hz (or 250~kbps/5000~kHz). 
Hence, OFDM1-MCS1 provides an increased robustness level while offering a higher spectral efficiency that enables to have denser deployments (a 333\% capacity increase). 

From these results, we can observe that the SINR required to provide PDR$>$80\% for OQPSK-DSSS largely depends on the packet length (4~dB difference).
In contrast, the SINR required to provide a PDR$>$80\% for the robust SUN-OFDM modulations remains constant ($<$1~dB) regardless of the packet length and the interference type.
This is owing to the fact that OQPSK-DSSS does not use any physical layer mechanism to enhance robustness except for the processing gain provided by DSSS.
Hence, the error probability increases as the packet length is increased.
For SUN-OFDM-based modulations, time and frequency symbol repetition reduces the probability of having an erroneous bit. 

\subsection{Recommendations}
\label{sec:recommendations}

Based on the results and the discussion presented above, we provide the following insights and recommendations to researchers and practitioners deploying low-power wireless communication systems using the SUN-OFDM physical layer in real-world environments:

\begin{enumerate}
    \item Thanks to the small effect of packet length in the PDR with respect to the SINR, SUN-OFDM allows to use larger packets (i.e.~packet bundling) to increase the transmission efficiency (i.e.~more effective data with the same packet headers) without sacrificing robustness. 
        In fact, SUN-OFDM allows payloads of up to 2047~bytes, effectively allowing to transmit full IPv6 packets without fragmentation or allowing to group up to sixteen 127-byte frames in a single 2047-byte frame.
    \item Despite the fact that SUN-OFDM transceivers consume a higher amount of energy compared to state-of-the-art IEEE~802.15.4 transceivers due to the additional circuitry required to operate (scrambler, convolutional encoder, puncturer, interleaver, Viterbi decoder), the higher level of robustness against interference provided by SUN-OFDM allows to use higher data rates (up to 800~kbps) to reduce the average energy consumption of the transmitter and the receiver devices.
    \item As the preamble of a SUN-OFDM packet is transmitted using the lowest MCS option of the current configuration, and includes information regarding the MCS option used to transmit the packet payload, this allows the transmitter to switch between different MCS options of the same SUN-OFDM configuration without any changes in the receiver configuration. 
        It is hence possible for the transmitter to use an aggressive modulation for an initial packet transmission, and use a more robust modulation when re-transmitting.
        Similarly, acknowledgement frames can be transmitted using the most robust modulation to increase the probability they are received.
    \item Use of SUN-OFDM for deployments with a high device density and/or high interference levels is advisable, as it provides higher spectral efficiency, while maintaining a similar level of robustness against interference with respect to OQPSK-DSSS.
        If interference is of concern, choosing OFDM1-MCS1 over OQPSK-DSSS translates into a 4x capacity increase while providing an average advantage of 9~dB against the same interference.
        In contrast, if interference is not a concern, choosing OFDM4-MCS5 offers a 26x capacity increase, while maintaining a similar level of protection against the same type of interference.
\end{enumerate}

\section{Conclusions}
\label{sec:conclusions}

This article evaluates the interference robustness of the OQPSK-DSSS and the SUN-OFDM physical layers defined in the IEEE~802.15.4-2015 standard.
Interfering conditions are generated in a controlled setup to evaluate the SINR required by each modulation to achieve PDR$>$80\%.
Compared to OQPSK-DSSS, results show that SUN-OFDM provides at least 6~dB additional protection regardless of interference type and packet length.
In addition, SUN-OFDM only occupies 1.2~MHz bandwidth, whereas OQPSK-DSSS occupies 5~MHz. 
This results in a higher spectral efficiency that allows one to have denser deployments or to trade bandwidth efficiency and interference robustness depending on the application requirements.
Overall, the presented results become a useful tool to understand the behavior of each particular modulation under interference conditions, and motivates the use of the SUN-OFDM physical layer to deploy low-power wireless systems in industrial scenarios.

\section*{Acknowledgment}
This research is partially supported by the SPOTS project (RTI2018-095438-A-I00) funded by the Spanish Ministry of Science, Innovation and Universities.

\bibliographystyle{IEEEtran}
\bibliography{bibliography}

\end{document}